\documentclass{article}
\usepackage{spconf,amsmath,graphicx,booktabs,multirow,makecell,caption,cite}
\usepackage{hyperref}
\hypersetup{
  colorlinks   = true, 
  urlcolor     = blue, 
  linkcolor    = black, 
  citecolor    = black 
}
\makeatletter
\newcommand{\printfnsymbol}[1]{%
  \textsuperscript{\@fnsymbol{#1}}%
}
\makeatother
\usepackage{wrapfig}
\usepackage{adjustbox}
\usepackage{float}
\usepackage{amssymb}
\usepackage{CJKutf8}
\usepackage{tablefootnote}
\usepackage{enumitem}
\usepackage{ragged2e}
\usepackage{microtype}                 
\newenvironment{justified_itemize}{%
    \begin{itemize}[rightmargin=0pt,topsep=0pt]
  \setlength{\itemsep}{0pt}%
  \setlength{\parskip}{0pt}%
}{%
  \end{itemize}
}
\usepackage[hang]{footmisc}
\makeatletter
\newcommand\footnoteref[1]{\protected@xdef\@thefnmark{\ref{#1}}\@footnotemark}
\makeatother
\newcommand{\chinese}[1]{\begin{CJK}{UTF8}{bsmi}#1\end{CJK}}
\title{BiSinger: Bilingual Singing Voice Synthesis }
\name{Huali Zhou\textsuperscript{1,2}\printfnsymbol{1}, Yueqian Lin\textsuperscript{2}\printfnsymbol{1}, Yao Shi\textsuperscript{2}, Peng Sun\textsuperscript{2}, Ming Li\textsuperscript{1,2}\printfnsymbol{2}\thanks{\scriptsize
\printfnsymbol{1} Equal contribution.\\ \printfnsymbol{2} Corresponding author. E-mail: \href{mailto:ming.li369@dukekunshan.edu.cn}{ming.li369@duke.edu}}}

\address{\textsuperscript{1}School of Computer Science, Wuhan University, Wuhan, China\\
\textsuperscript{2}Suzhou Municipal Key Laboratory of Multimodal Intelligent Systems, \\Duke Kunshan University, Kunshan, China}

\copyrightnotice{979-8-3503-0689-7/23/\$31.00~\copyright2023 IEEE}

\begin{document}

\maketitle
\begin{abstract}
Although Singing Voice Synthesis (SVS) has made great strides with Text-to-Speech (TTS) techniques, multilingual singing voice modeling remains relatively unexplored. This paper presents BiSinger, a bilingual pop SVS system for English and Chinese Mandarin. Current systems require separate models per language and cannot accurately represent both Chinese and English, hindering code-switch SVS. To address this gap, we design a shared representation between Chinese and English singing voices, achieved by using the CMU dictionary with mapping rules. We fuse monolingual singing datasets with open-source singing voice conversion techniques to generate bilingual singing voices while also exploring the potential use of bilingual speech data. Experiments affirm that our language-independent representation and incorporation of related datasets enable a single model with enhanced performance in English and code-switch SVS while maintaining Chinese song performance. Audio samples are available at \url{https://bisinger-svs.github.io}.
\end{abstract}

\vspace{-0.4cm}
\section{Introduction}
\label{sec:intro}
\vspace{-0.2cm}
Singing voice synthesis (SVS) is becoming increasingly popular in our daily lives. It aims to create natural and expressive singing voices that match the music score. SVS is a unique type of Text-to-Speech (TTS) task because it must strictly adhere to pitch and duration limitations in the scores. Additionally, obtaining the necessary training data for SVS is more challenging due to copyright restrictions and complex annotation requirements.

The development of the SVS system is in parallel with the progressive development of the TTS system. XiaoiceSing \cite{lu2020xiaoicesing} uses the Fastspeech \cite{ren2019fastspeech} network to generate high-quality singing voices, while ByteSing \cite{gu2021bytesing} employs the autoregressive Tacotron-like \cite{wang2017tacotron} structure as the acoustic model. With the emergence of the end-to-end TTS model, VITS \cite{kim2021conditional} and VISinger \cite{zhang2022visinger} were also proposed, which can effectively mitigate the two-stage mismatch problem in singing voice generation. Multiple open-source toolkits, e.g., Sinsy \cite{oura2010recent}, Muskits \cite{shi2022muskits}, and NNSVS \cite{yamamoto2023nnsvs}, etc., were released to boost the development of SVS research. The diffusion probabilistic-based model, DiffSinger \cite{liu2022diffsinger}, has recently demonstrated superior performance to alternative methods. However, these systems mainly target monolingual pop songs, assuming the input lyrics are from a single language. There has been little exploration into multilingual singing voice modeling, which is essential because mixed language lyrics are prevalent in real singing. Our work seeks to address this issue by focusing on Chinese-English bilingual singing voice synthesis.

To develop a bilingual SVS model, it would be ideal to use a bilingual singing corpus with detailed music annotations. Unfortunately, the Children’s Song Dataset \cite{choi2020children} is one of the few bilingual datasets available, consisting of Korean and English. Regrettably, there is currently no publicly available Chinese-English singing dataset, and collecting one would be time-consuming and challenging due to the need for bilingual singers and manual annotations.
To overcome this challenge, on the one hand, we propose using existing monolingual singing corpora, specifically M4Singer \cite{zhang2022m4singer} for Chinese and NUS-48E \cite{duan2013nus} for English, with a language-independent representation \cite{cai2023cross} to build our model. On the other hand, our research explores how to synthesize bilingual singing voices with the help of a bilingual speech dataset, DB-4 from Data Baker\footnote{\label{db4-ref}\url{https://www.data-baker.com/en}}. The highlights of our contributions are as follows: 

\begin{justified_itemize}
\item We study how Chinese and English singing voices can have a shared representation to learn similar pronunciations crossing languages with annotation adaption.  
\item Our proposed approach converts existing monolingual singing datasets with established singing voice conversion (SVC) techniques to create bilingual singing voices.  
\item Considering the rich resource of speech data, we also look into the possibility of developing bilingual singing voices using a bilingual speech database.
\end{justified_itemize}

The paper is structured as follows. Section \ref{sec:rel} covers the related works on SVS and multilingual speech synthesis, while Section \ref{sec:method} presents our methodology for bilingual SVS. Experiment setup and results can be found in Section \ref{sec:exper}. Finally, Section \ref{sec:conclu} provides the conclusion and future work.

\vspace{-0.48cm}
\section{Related Works}
\label{sec:rel}
\begin{figure*}[!htb]
\centering
\includegraphics[width=1.0\textwidth]{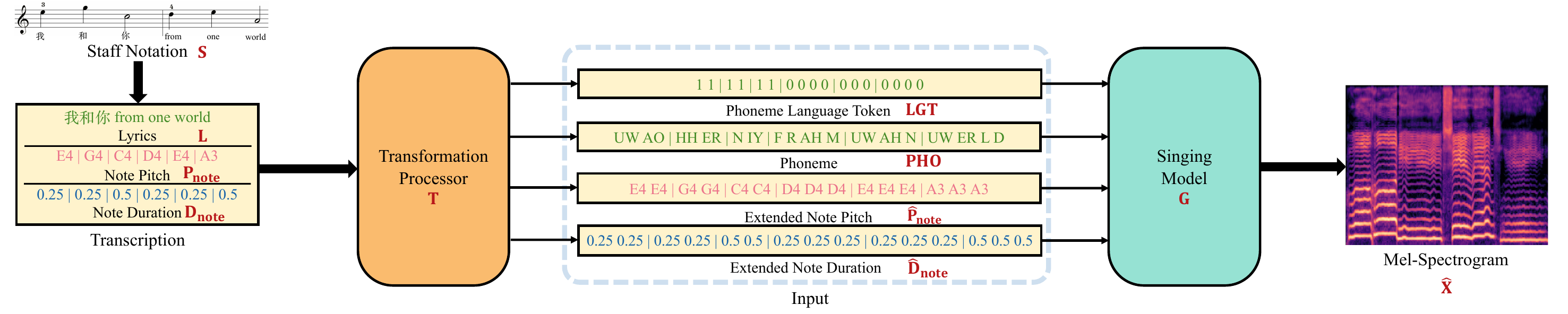}
\vspace{-0.36cm}
\caption{System overview.}
\label{fig:overview}
\end{figure*}
\vspace{-0.2cm}
\subsection{Singing Voice Synthesis}
\vspace{-0.1cm}
Singing voice synthesis (SVS) is a process that generates high-quality singing voices using music scores with lyrics for pronunciation and notes for prosody. Similar to the TTS task, it has evolved from the unit-selection method \cite{kenmochi2007vocaloid,bonada2016expressive}, which concatenates short waveform units from the database to more advanced statistical parametric systems like Hidden Markov Model (HMM)-based systems \cite{oura2010recent,nakamura2014hmm}, and now to mainstream deep neural network (DNN)-based architectures \cite{lu2020xiaoicesing,gu2021bytesing,zhang2022visinger,shi2022muskits,yamamoto2023nnsvs,liu2022diffsinger}. SVS typically uses a two-stage pipeline, consistent with the TTS task, where the acoustic model learns to map music score inputs to acoustic features, and the vocoder reconstructs the waveform based on the predicted acoustic feature. 

Some work involves extending existing singing voice synthesis systems to accommodate other languages. In \cite{nakamura2014hmm}, they adapted the HMM-based method for Japanese SVS to work with English by developing language-independent contexts. They also investigated new syllable allocation and duplication methods, considering the distinct syllable structures of Japanese and English. This work was aided by data from a bilingual singer, which provided informative insights, but unfortunately, the available data was quite scarce \cite{choi2020children}. With multi-singer Mandarin, English, and Cantonese singing data mined from music websites, \cite{ren2020deepsinger} could perform cross-lingual singing voice synthesis as a byproduct of multilingual training. However, it should be noted that systems in \cite{ren2020deepsinger} assume the inputs are in a single language. For existing systems, such as Muskits \cite{shi2022muskits}, multilingual synthesis requires separate models for different languages, making it incapable of synthesizing code-switch singing voices within a single model. 
\vspace{-0.2cm}
\subsection{Multilingual Speech Synthesis}
\vspace{-0.1cm}

Multilingual speech synthesis is designed to alleviate the need for training separate models for different languages and support low-resource synthesis. In order to promote the knowledge-sharing capacity of the model for different languages, a unified input representation \cite{zhang2019learning} of various languages has been sought for a long time. In \cite{li2019bytes}, the Unicode bytes were used across languages owing to their language independence and fixed size. Additionally, \cite{cao2019end} explored both a shared multilingual encoder with language embedding and a separate monolingual encoder. Another approach \cite{nekvinda2020one} involved using an additional network conditioned on language to generate parameters for multiple language-dependent encoders to enable cross-lingual knowledge-sharing. Cross-lingual synthesis was explored in \cite{cai2023cross}, proposing common phonemic representations linked to numeric language ID codes, which provides substantial inspiration to our research. 

To preserve different languages' characteristics while sharing knowledge with shared phoneme inputs, \cite{liu2019cross, liu2020multi} introduced stress and tone embedding. Experimental observations in \cite{lee2018learning} suggested that there is some degree of shared pronunciation across languages, which was helpful in low-resource scenarios. However, the close pronunciation between Chinese and English could lead to mutual interference, such as intonation variation and mispronunciation. To address this issue, \cite{yang2022improve} proposed an embedding strength modulator to capture the dynamic strength of language and phonology, which has also been incorporated into our work.

\section{Methodology}
\label{sec:method}
\vspace{-0.2cm}
\subsection{Formulation of the Model}
\vspace{-0.1cm}

Fig. \ref{fig:overview} presents an overview of our system. Given a staff notation $\mathbf{S}$, our first step is to transcribe it into lyrics $\mathbf{L}$, note pitch $\mathbf{P_{note}}$, and note duration $\mathbf{D_{note}}$. Subsequently, these transcribed elements are processed through a transformation processor $\mathbf{T}$ to format the data for the singing model $\mathbf{G}$. The detailed transformation process is illustrated as follows:

\vspace{-0.3cm}
\begin{equation}
\mathbf{LGT}, \mathbf{PHO}, \mathbf{\hat{P}_{note}}, \mathbf{\hat{D}_{note}} = \mathbf{T}(\mathbf{L}, \mathbf{P_{note}}, \mathbf{D_{note}})
\label{eq:trans}
\end{equation}
where $\mathbf{LGT}$ and $\mathbf{PHO}$ are the language tokens and universal phoneme representations generated according to the lyrics, $\mathbf{\hat{P}_{note}}$ is the phoneme's corresponding note pitch that is extended for phoneme-level alignment, and similarly, $\mathbf{\hat{D}_{note}}$ is the repeated input note duration.

After obtaining phoneme-level data, we then feed it into end-to-end singing model $\mathbf{G}$ to predict the Mel-spectrogram $\hat{\mathbf{X}}$, as shown in Eq. \ref{eq:model}, and calculate the loss according to $\mathbf{X}$, continually updating itself throughout the training stage.

\vspace{-0.2cm}
\begin{equation}
\hat{\mathbf{X}} = \mathbf{G}(\mathbf{LGT}, \mathbf{PHO}, \mathbf{\hat{P}_{note}}, \mathbf{\hat{D}_{note}})
\label{eq:model}
\end{equation}
\vspace{-0.2cm}

Once trained, the model has the capability to predict the Mel-spectrogram $\widetilde{\mathbf{X}}$ for any input music score, such as multilingual lyrics $\mathbf{L_{mul}}$, note pitch data $\mathbf{P_{note}}$, and note duration data $\mathbf{D_{note}}$. This prediction can be expressed by Eq. \ref{eq:pred}:

\vspace{-0.2cm}
\begin{equation}
\widetilde{\mathbf{X}} = \mathbf{G}(\mathbf{T}(\mathbf{L_{mul}}, \mathbf{P_{note}}, \mathbf{D_{note}}))
\label{eq:pred}
\end{equation}
\vspace{-0.8cm}
\subsection{Language-independent Representation}
\vspace{-0.1cm}

Inspired by \cite{cai2023cross}, we adopt the CMU Pronunciation  Dictionary\footnote{\label{cmudict}\url{http://www.speech.cs.cmu.edu/cgi-bin/cmudict}}
as the shared phoneme representation for  Chinese and English languages to overcome the challenge of different grapheme or phoneme sets in multilingual SVS. For Mandarin, characters are first represented as Pinyin using Pypinyin\footnote{\label{pypinyin}\url{https://github.com/mozillazg/python-Pinyin}}, which can be converted to CMU phonemes by the Pinyin-to-CMU mapping table\footnote{\label{pinyin2cmu}\url{https://github.com/kaldi-asr/kaldi/blob/master/egs/hkust/s5/conf/pinyin2cmu}}.
Likewise, each English word can be converted into CMU phonemes by referring to the mapping table for English, with examples in Table \ref{tab:phoneme-examples}. When it comes to singing, the pitch is primarily determined by the score rather than tone and accent, according to \cite{gu2021bytesing}. Therefore, tone or stress information is not taken into consideration. Furthermore, the code-switched song shares the same BPM across linguistic boundaries.

\begin{table}[h]
\caption{Unit-to-CMU phoneme mapping examples.}
\centering
\label{tab:phoneme-examples}
\resizebox{0.35\textwidth}{!}{
\begin{tabular}{@{}cc|cc@{}}
\toprule
\thead{Chinese Pinyin} & \thead{CMU Phonemes} & \thead{English Word} & \thead{CMU Phonemes} \\ 
\midrule
rang  & R AE NG   & cat      & K AE T    \\
wo    & W AO      & fan      & F AE N    \\
nuan  & N UW AE N & song     & S AO NG   \\
yang  & Y AE NG   & total    & T OW T AH L  \\
zhui  & JH UW IY  & story    & S T AO R IY  \\
\bottomrule
\end{tabular}
}
\end{table}

It is worth noting that the pronunciation of the same phoneme can differ between Chinese and English. Due to this, we use language identification tokens to preserve language-dependent characteristics while sharing unified phonemes. The tokens `0' and `1' represent the language ID for each phoneme, with `0' signifying English and `1' indicating Chinese. Take the lyrics ``\chinese{我和你} from one world" as an example; two token sequences are obtained as language-independent representations. The first is the phoneme sequence ``UW AO HH ER N IY F R AH M UW AH N UW ER L D," and the second is the corresponding language token sequence ``1 1 1 1 1 1 0 0 0 0 0 0 0 0 0 0 0," which has the same length as the former.

\vspace{-0.2cm}
\subsection{Language-style-infused Encoder}
\label{sec:infused-encoder}
\vspace{-0.1cm}

\begin{figure}[htb]
\centering
\includegraphics[width=0.47\textwidth]{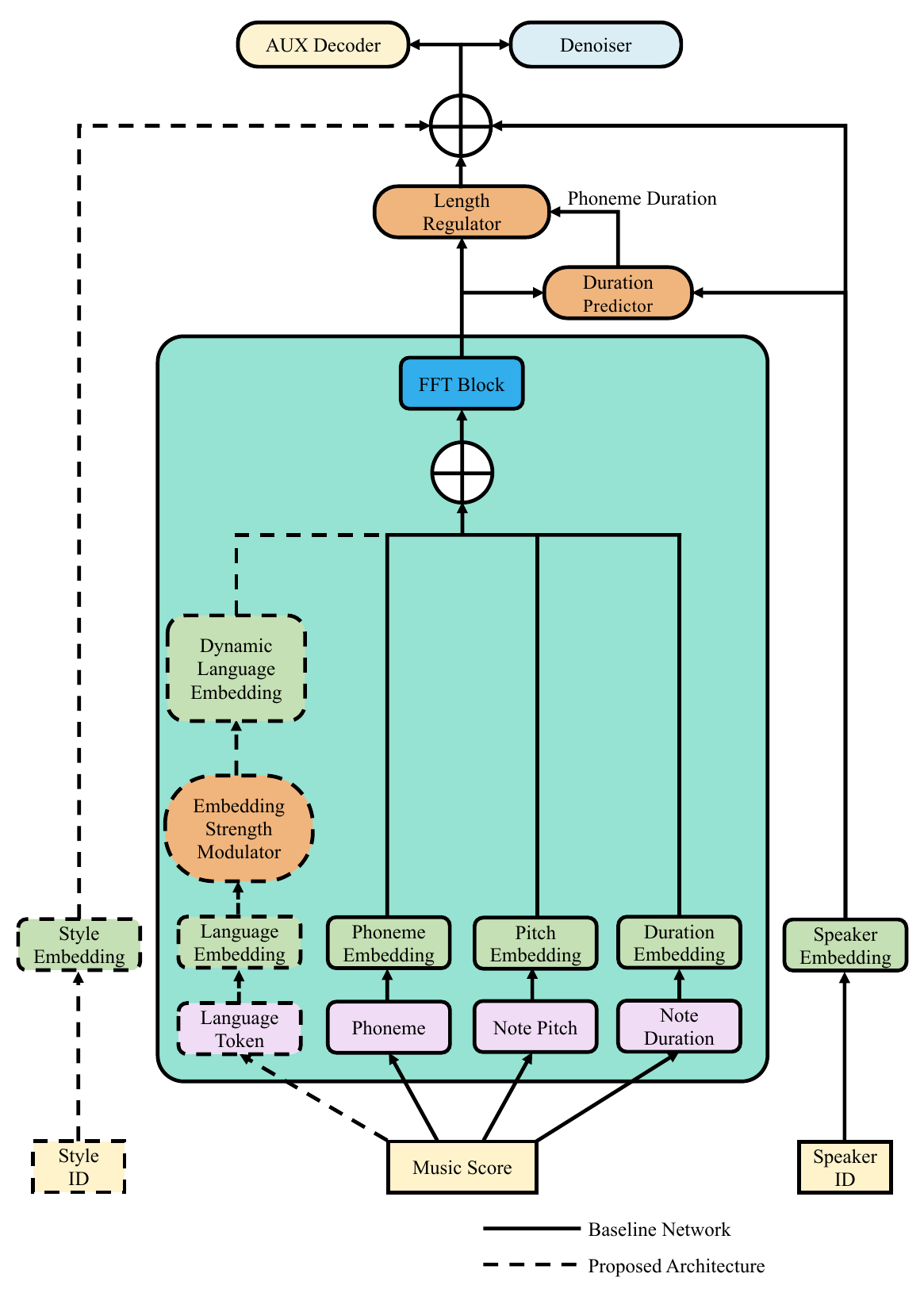}
    \caption{Language-style-infused encoder.}
    \label{fig:encode}
\end{figure}

We modify the encoder in DiffSinger \cite{liu2022diffsinger} to suit our multilingual singing voice modeling with both speech and singing data, as illustrated in Fig. \ref{fig:encode}.

To deal with the multilingual issue, a learnable language embedding layer is introduced to convert the numeric language ID tokens to a 256-dimensional language embedding sequence, then encoded with the original phoneme embedding, note embedding and note duration embedding. Additionally, we employ an Embedding Strength Modulator (ESM) \cite{yang2022improve}, a fusion of multi-head attention scheme and feed-forward network, to capture the dynamic strength of phonology and language, as shown in Fig. \ref{fig:encode}. When utilizing the ESM module, the feed-forward Transformer (FFT) block employs the dynamic language embedding that incorporates the encoded phoneme embedding instead of the static data directly calculated based on the language ID.

A style identification token for each utterance is introduced to address the training imbalance between speech and singing data and avoid the synthesized singing voices from sounding excessively smooth and fast like speech. This token consists of the numbers `0', `1', and `2', with `0' indicating speech, `1' indicating singing, and `2' indicating pseudo-singing obtained through the pitch shift method illustrated in Sec \ref{sec:db4-shift}. As shown in Fig. \ref{fig:encode}, a style embedding layer is designed, converting the numeric style ID token to a 256-dimensional style embedding. Lastly, we merge the style embedding, speaker embedding, and the result of the encoded musical score to form the input for the auxiliary decoder or the denoiser.

\vspace{-0.2cm}
\subsection{Adaptation of Datasets}
\vspace{-0.1cm}

As mentioned in Section \ref{sec:rel}, the scarcity of well-annotated singing data poses crucial challenges for SVS, especially multilingual ones. We propose three methods to address this challenge: phoneme-level musical annotation adaption, timbre conversion, and pseudo-singing through pitch shift.

\vspace{-0.2cm}
\subsubsection{Phoneme-level Musical Annotation Adaptation}
\label{sec:m4-adaption}
\vspace{-0.1cm}

To fully utilize the existing datasets' annotation information, we intend to adjust the original annotations to fit our CMU phoneme set. The note and note durations should be considered thoughtfully for singing voice datasets. We propose a straightforward approach that involves splitting each phoneme in the Pinyin set into its corresponding CMU phonemes; for instance, `ang' will be split into `AE NG.'
For the musical annotations, based on the mapping relationship, we duplicate the note and note duration from the original annotation for the corresponding CMU phonemes. Next, we divide the original phoneme duration equally among the split CMU phonemes and assign the average duration to each CMU phoneme.

Moreover, we noticed that simply averaging phoneme duration may pose issues since vowels in singing tend to be longer than consonants. The average strategy can negatively affect the naturalness of synthesized singing voices. To handle this, we present distributing phoneme duration proportionally based on the Montreal Forced Aligner (MFA) \cite{mcauliffe2017montreal} coarse alignment. Specifically, we utilize MFA\footnote{We train separate MFA models for M4Singer, Chinese speech data in DB-4, and English speech data in DB-4, using complete datasets of 29.77h, 11.84h and 5.68h, respectively.} to align with the CMU dictionary and obtain phoneme duration. However, the alignment results are not accurate enough when it comes to the long duration of vowels and portamento. We observed that the phoneme duration closely matches M4Singer's original annotations for the initial components of Pinyin. The ratio between consonants and vowels appears reasonable for the final components. Bearing these observations in mind, we retain the same phoneme duration for Pinyin initials when converting them to CMU-based phonemes. For the Pinyin finals, we first convert them into CMU phonemes and then proportionally distribute the original duration based on the ratio obtained in the MFA alignment result. 
Lastly, we calculate each mapped CMU phoneme's note and note duration based on adapted phoneme durations and original note annotations. A concrete example can be found in Table \ref{tab:three annotation}.

\begin{table}[h]
  \centering
  \caption{Three annotation types.}
  \vspace{-0.3cm}
  \label{tab:three annotation}
\resizebox{0.29\textwidth}{!}{
    \begin{tabular}{cccccc}
      \toprule
      \multicolumn{2}{c}{\text{original}} & \multicolumn{2}{c}{\text{average}} & \multicolumn{2}{c}{\text{proportional}} \\
      \cmidrule(r){1-2} \cmidrule(lr){3-4} \cmidrule(l){5-6}
      \thead{phs} & \thead{ph\_dur} & \thead{phs} & \thead{ph\_dur} & \thead{phs} & \thead{ph\_dur} \\
      \midrule
      c   & 0.18  & T  & 0.09   & T  & 0.036 \\
      uen & 0.245 & S  & 0.09   & S  & 0.144 \\
      uen & 0.295 & UW & 0.0817 & UW & 0.18  \\
          &       & AH & 0.0817 & AH & 0.065 \\
          &       & N  & 0.0817 & AH & 0.115 \\
          &       & UW & 0.0983 & N  & 0.18  \\
          &       & AH & 0.0983 &    &       \\
          &       & N  & 0.0983 &    &       \\
      \bottomrule
    \end{tabular}
}
\end{table}

\vspace{-0.5cm}
\subsubsection{Timbre Conversion Method}
\label{sec:nus-svc}
\vspace{-0.1cm}

While the shared CMU phonemes can help our model learn English pronunciation from Chinese data to a degree, some phonemes, such as /TH/, /Y/, /IH/, /DH/, /V/, and /OY/, are absent in Chinese. Substituting these missing phonemes with phonetically similar ones results in less natural-sounding voices, reinforcing the necessity of English data for a comprehensive and accurate representation of English phoneme pronunciation.

Due to the limited scale of existing English singing voice datasets and the need for consistent timbre to accurately capture how phonemes are pronounced in both English and Chinese, we have come up with a new method that uses advanced SVC techniques, so-vits-svc\footnote{\label{so-vits-svc}\url{https://github.com/svc-develop-team/so-vits-svc}}, an open-source repository for robust singing voice conversion. Our approach involves the conversion of original singing voices from a small English singing dataset to match all the different timbres in a larger Chinese singing dataset, significantly expanding the English singing voice dataset while creating a smooth transition between both languages.

Table \ref{tab:adjust} illustrates our pitch adjustment methodology when converting between different voice types, namely: Bass, Baritone, Tenor, Alto, and Soprano. Our goal is to preserve the natural voice range of the target singer and reduce conversion artifacts. To achieve this, we adjust the pitch in semitones. For instance, conversion from Bass to Tenor incurs a pitch shift of +8 semitones, increasing the pitch of the original Bass voice to match the typical Tenor range. 

\begin{table}[h]
\centering
\caption{Pitch adjustments reference (in semitones).}
  \vspace{-0.3cm}
\resizebox{0.35\textwidth}{!}{
\begin{tabular}{cccccc}
\toprule
\multicolumn{1}{c}{\multirow{2}{*}{From}} & \multicolumn{5}{c}{To} \\
\cmidrule(lr){2-6}
& Bass & Baritone & Tenor & Alto & Soprano \\
\midrule
Bass & 0 & +4 & +8 & +12 & +12 \\
Baritone & -4 & 0 & +4 & +8 & +8 \\
Tenor & -8 & -4 & 0 & +4 & +8 \\
Alto & -12 & -8 & -4 & 0 & +4 \\
Soprano & -12 & -8 & -8 & -4 & 0 \\
\bottomrule
\end{tabular}
}
\label{tab:adjust}
\end{table}
\vspace{-0.5cm}
\subsubsection{Musical Annotation and Pitch Shift}
\label{sec:db4-shift}
\vspace{-0.1cm}

With easy access to existing large-scale, high-quality speech corpora, we plan to explore using a bilingual speech corpus for the bilingual SVS task. We treat speech data as plain singing data, although it lacks some singing characteristics. Unfortunately,  there is no musical annotation for speech corpus, which is essential for SVS. Likewise, we utilize MFA for speech alignment, with CMU phoneme durations and word boundaries as results. We then extract F0 using the open-source Parselmouth\footnote{\url{https://github.com/YannickJadoul/Parselmouth}}, which is then averaged according to word boundaries to get musical notes. We can also calculate phoneme duration and note duration from the alignment results.

To ensure the synthesized singing voice is as rhythmic as possible, we adjust the pitch of speech data with the WORLD vocoder \cite{morise2016world} to achieve data augmentation and prevent the synthesized voice from being too stable and lacking musical expressiveness. To clarify, we start by defining ten frequently used melody frequencies. With the help of WORLD, we extract pitch contour (F0), aperiodic spectral envelope (AP), and harmonic spectral envelope (SP) for each speech audio. Next, we randomly choose one melody and replace the original F0 with it while keeping the F0 contour length unchanged. We then use WORLD to synthesize a new pseudo-singing voice based on the shifted F0 and the original AP and SP. Lastly, we update the note information in the musical annotation to match the corresponding melody. This way, we obtain pseudo-singing voices rich in pitch variation from the original stationary speech data.

\section{Experiments}
\label{sec:exper}
\vspace{-0.3cm}
\subsection{Datasets}
\vspace{-0.1cm}
We use three datasets for experiments, namely, M4Singer \cite{zhang2022m4singer}, NUS-48E \cite{duan2013nus} and DB-4\footnoteref{db4-ref},  as summarized in Table \ref{tab:dataset}. 
Singing and speech audios are recorded at 44.1kHz, and 48kHz, respectively, with 16-bit quantization, and we downsample them to 24kHz in our experiments.

\begin{table}[h]
    \centering
    \caption{Datasets used in our experiments.}
    \vspace{-0.3cm}
    \resizebox{0.49\textwidth}{!}{\begin{tabular}{cccccccc} \toprule
        \multirow{2.5}{*}{Dataset} & \multirow{2.5}{*}{Language} & \multirow{2.5}{*}{Voice parts} & \multirow{2.5}{*}{\shortstack{Duration\\(hour)}}  & \multicolumn{2}{c}{\# speakers} & \multicolumn{2}{c}{\# utterances} \\
        \cmidrule(lr){5-6} \cmidrule(lr){7-8}
        & & & & \multicolumn{1}{c}{M} & \multicolumn{1}{c}{F} & \multicolumn{1}{c}{Singing} & \multicolumn{1}{c}{Speech} \\
        \midrule
        M4Singer \cite{zhang2022m4singer} & CN & S, A, T, $\text{B}_{1}$\tablefootnote{\label{B12}$\text{B}_{1}$, the abbreviation of Bass; $\text{B}_{2}$, the abbreviation of Baritone} & 29.77 & 10 & 10 & 20942 & --\\
        NUS-48E \cite{duan2013nus} & EN & S, A, T, $\text{B}_{1}$, $\text{B}_{2}$\footnoteref{B12} & 1.91 & 6 & 6 & 1262 & --\\
        \multirow{2}{*}{DB-4}\footnoteref{db4-ref} & CN & \multirow{2}{*}{--} & 11.84 & \multirow{2}{*}{0} & \multirow{2}{*}{1} & \multirow{2}{*}{--} & 10000 \\ 
        &EN & & 5.68 & & & & 5000 \\
        \bottomrule
    \end{tabular}}
    \label{tab:dataset}
\end{table}

To be detailed, M4Singer \cite{zhang2022m4singer} are aligned on the Pinyin-phoneme-level, which cannot capture English pronunciation. An example from M4Singer is provided in Table \ref{tab:m4singer-examples}. Phonemes with corresponding musical notes and note durations serve as input to the acoustic model's front end, carrying phoneme durations for accurate phoneme prediction.
\vspace{-2pt}
\begin{table}[!htb]
\caption{Phoneme-level annotation in M4Singer.}
\vspace{-0.2cm}
\centering
\label{tab:m4singer-examples}
\vspace{-0.1cm}
\resizebox{0.24\textwidth}{!}{
    \begin{tabular}{@{}ccccc@{}}
        \toprule
        \thead{phs} & \thead{is\_slur} & \thead{ph\_dur} & \thead{notes} & \thead{notes\_dur}\\
        \midrule
        r & 0 & 0.28 & 56 & 0.63 \\
        ang & 0 & 0.35 & 56 & 0.63 \\
        uo & 0 & 0.23 & 56 & 0.23 \\
        n & 0 & 0.32 & 58 & 0.51 \\
        uan & 0 & 0.19 & 58 & 0.51 \\
        iang & 0 & 0.0905 & 61 & 0.0905 \\
        iang & 1 & 0.1636 & 60 & 0.1636 \\
        iang & 0 & 0.9 & 58 & 0.9 \\
        \bottomrule
    \end{tabular}
    }
\end{table}
\vspace{-0.7cm}
\subsection{Experimental Setup}
\vspace{-0.1cm}

To verify the proposed methods, we conduct experiments based on the DiffSinger model \cite{liu2022diffsinger} and settings \cite{zhang2022m4singer}, using the pretrained HiFi-GAN\cite{kong2020hifi} provided by M4Singer\cite{zhang2022m4singer}
as vocoder. There are two types of model structures and three corpora, and systems to compare are described in Table \ref{tab:systems}.
\begin{justified_itemize}
\item \textbf{Model 1}: Original DiffSinger model.
\item \textbf{Model 2}: DiffSinger with Language-style-infused encoder illustrated in Sec \ref{sec:infused-encoder}.
\item \textbf{Corpus 1}: M4Singer with averaged CMU-phoneme-based annotation described in Sec \ref{sec:m4-adaption}, totally $29.77h$.
\item \textbf{Corpus 2}: M4Singer with proportional CMU-phoneme-based annotation described in Sec \ref{sec:m4-adaption}, totally $29.77h$.
\item \textbf{Corpus 3}: As described in Sec \ref{sec:nus-svc}, we use so-vits-svc\footnoteref{so-vits-svc} to convert the singing voices in the NUS-48E dataset to the target 20 singers in the M4Singer dataset, thereby each singer owns both Mandarin and English singing data, with a total of $1.91h \times 20=38.2h$.
\item \textbf{Corpus 4}: As stated in Sec \ref{sec:db4-shift}, we get speech DB-4 and pseudo-singing DB-4, $11.84h+5.68h=17.52h$ each.
\end{justified_itemize}

\vspace{-0.16cm}
\begin{table}[!htb]
\caption{System description.}
\vspace{-0.24cm}

\centering
\label{tab:systems}
\resizebox{0.49\textwidth}{!}
{
\begin{tabular}{@{}cccccc@{}}
\toprule
\textbf{System} & \textbf{System 1} & \textbf{System 2} & \textbf{System 3} & \textbf{System 4} & \textbf{System 5} \\ 
\midrule
\textbf{Model} & Model 1 & Model 1 & Model 2 & Model 2 & Model 2 \\
\textbf{Corpora} & Corpus 1 & Corpus 2 & Corpus 2, 3 & Corpus 2, 4 & Corpus 2, 3, 4 \\
\bottomrule
\end{tabular}
}
\end{table}

\vspace{-0.3cm}
\subsection{Evaluation}
\vspace{-0.2cm}

We create 25 test cases for each language scenario, cumulating a total of 75 cases. To compare the systems\footnote{Please note that when encountering missing phonemes, we replace them with phonetically similar ones for \textbf{System 1} and \textbf{System 2}.}, we conduct both objective and subjective evaluations with four target speakers, each of whom represents a different voice part. We recruit a qualified singer to record clean singing voices for each test case, then converted the recording to the four target timbres using so-vits-svc\footnoteref{so-vits-svc} as ground truth.

For the objective evaluation, following \cite{shi2022muskits}, we utilize four metrics, including Melcepstrum distortion (MCD), F0 root mean square error (F0\_RMSE), voice/unvoiced error rate (VUV{\_}E), and semitone accuracy (SA). We employ Whisper \cite{radford2023robust} to compute the word error rate (WER), which serves as our multilingual evaluation criterion. 
We use a ResNet101-based speaker verification model, trained with the VoxCeleb2 development set and achieving an Equal Error Rate (EER) of 0.44\% on the Vox-O test set, to extract speaker embeddings and calculate cosine similarity (SIM) for the synthesized singing, reaching around 0.6 for the three voice parts: Tenor, Alto, and Soprano, which objectively exhibits the speaker’s identity well hold\footnote{The highly-precise verification model results in a relatively strict score.}.
For the subjective evaluation, 14 native Chinese speakers with work-proficient English are recruited to score the entire set of generated samples, ranging from 1 to 5 with 0.5 increments. 
The results are shown in Table \ref{table:MOS_comparison}. 

\begin{figure}[htb]
    \centering
    \includegraphics[width=0.41\textwidth]{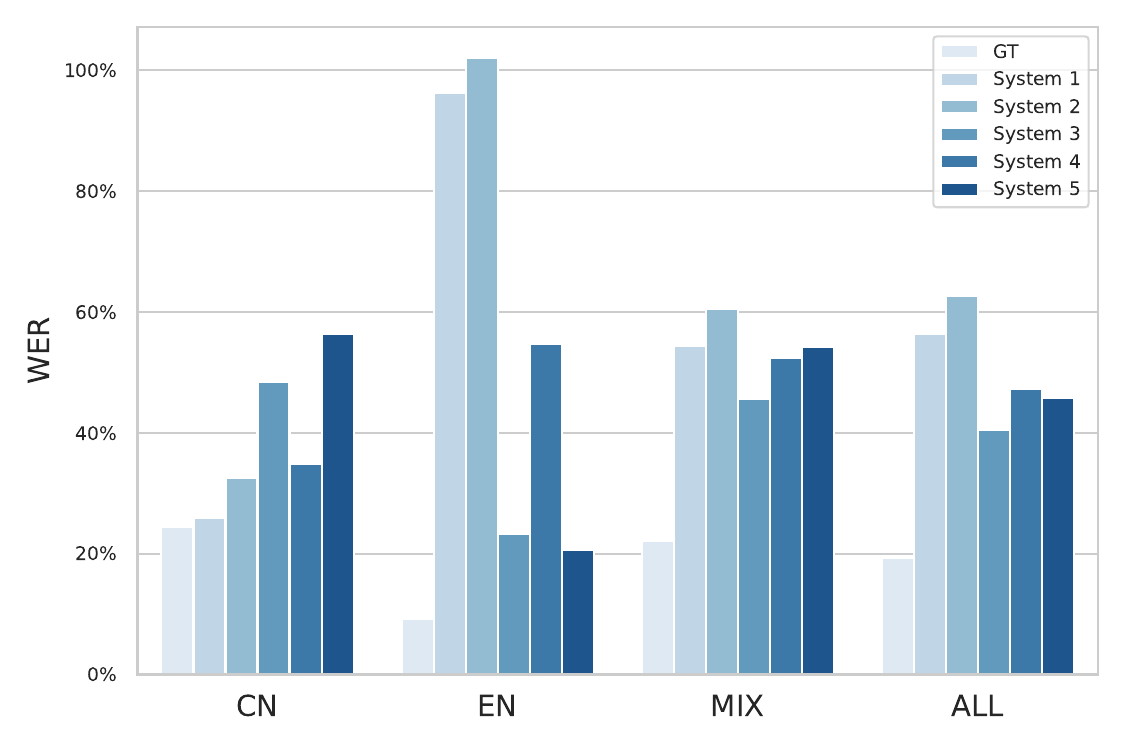}
    \vspace{-0.5cm}
    \caption{WER results.}
    \label{fig:wer}
\end{figure}

\begin{table}[!htb]
    \centering
    \caption{Experimental results in terms of objective and subjective metrics.}
    \vspace{-0.2cm}
    \resizebox{0.49\textwidth}{!}
    {
    \begin{tabular}{lccccccc}
    \toprule 
    System & Language & MCD $\downarrow$ & F0\_RMSE $\downarrow$ & VUV\_E $\downarrow$ & SA $\uparrow$ & SIM $\uparrow$ & MOS\tablefootnote{\label{mos_ci}With 95\% confidence interval.} $\uparrow$ \\ 
    \midrule
\multirow{4}{*}{\bf{Ground Truth}}  
& CN    & \multirow{4}{*}{--} & \multirow{4}{*}{--} & \multirow{4}{*}{--}  & \multirow{4}{*}{--} & {--} &  $3.78 \pm 0.08$ \\
& EN    & & & & & {--} &  $3.80 \pm 0.08$ \\
& MIX   & & & & & {--} &  $3.60 \pm 0.09$ \\
& ALL   & & & & & {--} &  $3.73 \pm 0.05$ \\
\midrule
\multirow{4}{*}{\bf{System 1}}  
& CN    & \textbf{9.5983} & \textbf{0.1607} & \textbf{6.53\%} & \textbf{45.16\%} & 0.64 & $\mathbf{3.41 \pm 0.09}$ \\
& EN    & 10.1743 & 0.1654 & \textbf{6.36\%} & 39.95\% & 0.59 & $2.75 \pm 0.09$ \\
& MIX   & 10.8097 & 0.2312 & \textbf{9.01\%} & 34.67\% & 0.65 & $3.06 \pm 0.09$ \\
& ALL   & 10.1941 & 0.1858 & \textbf{7.30\%} & 39.93\% & 0.63 & $3.07 \pm 0.05$  \\
\midrule
\multirow{4}{*}{\bf{System 2}} 
& CN    & 9.7394  & 0.1810 & 7.17\% & 43.18\% & 0.62 &$3.34 \pm 0.09$ \\
& EN    & 10.2490 & 0.1883 & 8.40\% & 36.25\% & 0.57 &$2.56 \pm 0.09$ \\
& MIX   & 11.0579 & 0.2293 & 9.93\% & 33.51\% & 0.62 &$2.90 \pm 0.09$ \\
& ALL   & 10.3488 & 0.1995 & 8.50\% & 37.64\% & 0.60 &$2.96 \pm 0.05$ \\
\midrule
\multirow{4}{*}{\bf{System 3}} 
& CN    & 9.7953  & 0.1709 & 7.29\% & 43.38\% & 0.58 &$2.97 \pm 0.09$ \\
& EN    & \textbf{8.6386} & \textbf{0.1440} & 6.93\%  & \textbf{45.05\%} & 0.52 &$3.40 \pm 0.08$ \\
& MIX   & 10.3429 & \textbf{0.2163} & 9.69\% & \textbf{38.88\%} & 0.58 & $\mathbf{3.32 \pm 0.08}$ \\
& ALL   & \textbf{9.5923} & \textbf{0.1770} & 7.97\%  & \textbf{42.44\%} & 0.56 & $3.23 \pm 0.05$ \\
\midrule
\multirow{4}{*}{\bf{System 4}} 
& CN    & 9.8097  & 0.1711 & 7.00\% & 44.14\% & 0.60 & $3.14 \pm 0.09$ \\
& EN    & 10.4531 & 0.1854 & 7.69\% & 33.11\% & 0.49 & $2.58 \pm 0.09$ \\
& MIX   & 11.3648 & 0.2262 & 10.54\% & 31.95\% & 0.58 & $2.89 \pm 0.08$ \\
& ALL   & 10.5425 & 0.1942 & 8.41\% & 36.40\% & 0.56 & $2.89 \pm 0.05$ \\
\midrule
\multirow{4}{*}{\bf{System 5}} 
& CN    & 9.8542 & 0.1761 & 7.15\% & 44.28\% & 0.57 &$3.04 \pm 0.09$ \\
& EN    & 8.7834 & 0.1541 & 7.68\% & 41.93\% & 0.49 & $\mathbf{3.42 \pm 0.09}$ \\
& MIX   & \textbf{10.2845} & 0.2182 & 10.14\% & 37.21\% & 0.59 & $3.24 \pm 0.09$ \\
& ALL   & 9.6407 & 0.1828  & 8.33\% &  41.14\% & 0.55 & $\mathbf{3.24 \pm 0.05}$ \\
\bottomrule
\end{tabular}}
\label{table:MOS_comparison}
\end{table}
Based on the WER (Fig. \ref{fig:wer}) and MOS results, it is evident that \textbf{System 1} and \textbf{System 2} perform poorly on English songs. With the integration of Corpus 3, \textbf{System 3} and \textbf{System 5} have shown significant improvement, with MOS scores of 3.40 and 3.42, respectively, compared to \textbf{System 2}'s score of 2.56. 
Comparing the results from \textbf{System 3, 4, 5}, using Corpus 4 has effectively alleviated the decline in the Chinese synthesis effect. At the same time, the real English singing data in Corpus 3 has assisted the model in effectively learning English from Corpus 4, consisting of speech data and pseudo-singing data.

\begin{figure}[htb]
    \centering
    \includegraphics[width=0.46\textwidth]{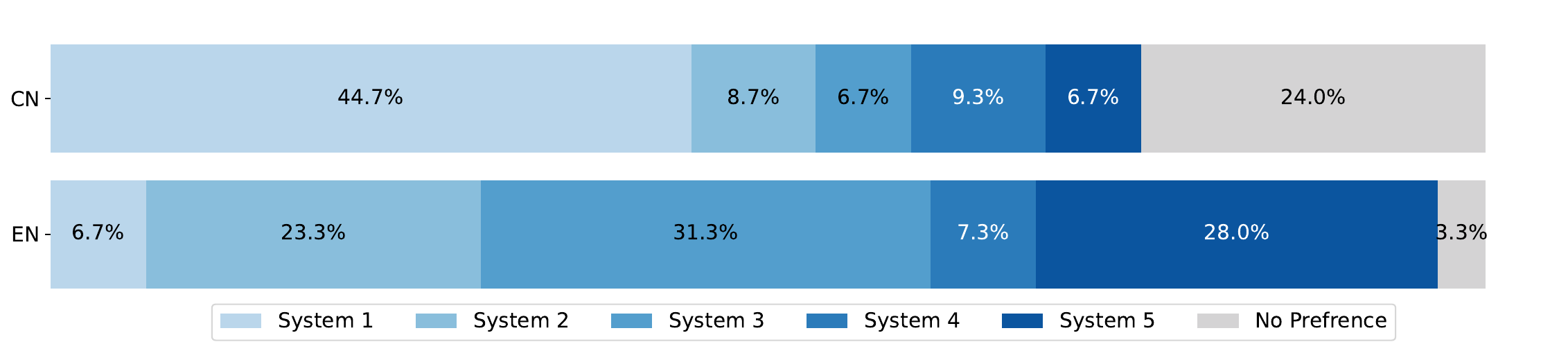}
    \vspace{-0.2cm}
    \caption{Preference results in terms of pronunciation.}
    \label{fig:pre}
\end{figure}

To compare these systems more effectively, we conducted a preference test, asking the subjects to choose the best system in pronunciation (Fig. \ref{fig:pre}) for Chinese and English songs. We found that \textbf{System 1} performed best in Chinese; however, its understanding of Chinese phoneme duration proportions could not be transferred to English synthesis. With the proportionally distributed phoneme duration, the synthesized effect of Chinese is still maintained, as evidenced by the results of \textbf{System 2, 3, 4, 5}. Furthermore, the proportionally obtained duration proved more suitable for English tasks, compared to \textbf{System 1}. It is worth noting that even though \textbf{System 2}, trained exclusively with Chinese data, could generate English singing voices pleasing to the human ear, the large model, Whisper\cite{radford2023robust}, classified such pronunciation effects as Chinese based on its transcription results and WER calculations. However, after adding Corpus 3 and Corpus 4, the English WER results of \textbf{System 3, 4, 5} have dropped significantly, indicating that real English data is necessary to learn English phonemes' pronunciation better. Particularly, according to preference results, \textbf{System 3} and \textbf{System 5} have gained substantial benefits from incorporating real English singing data.
\vspace{0.1cm}
\vspace{-0.5cm}
\subsection{Discussion on Phoneme Substitution}
\vspace{-0.1cm}

\begin{figure}[!htb]
    \centering
    \includegraphics[width=0.49\textwidth]{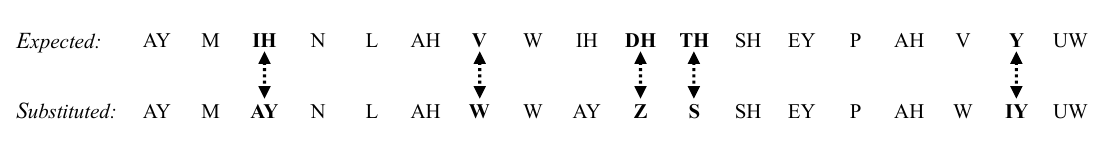}
    \vspace{-0.7cm}
    \caption{Substitution example.}
    \label{fig:substitute}
\end{figure}

Based on the results from \textbf{System 1} and \textbf{System 2}, it is evident that shared CMU phonemes can partially assist our model in learning English pronunciation from Chinese data. Most of these phonemes are present in both languages. However, certain English phonemes like /TH/, /Y/, /IH/, /DH/, /V/, and /OY/ are absent in Chinese data. During the inference process, these are substituted with similarly pronounced Chinese phonemes. However, this approach may only sometimes work since some phonemes lack similar pronunciations in Chinese. In these situations, the pronunciation of each phoneme remains isolated and cannot be connected coherently as a word. Frequent substitutions may also result in synthesized voices that sound unclear and similar to Chinese. Consider the lyrics ``I'm in love with the shape of you'' as an example (Fig. \ref{fig:substitute}); the system with phoneme substitution performs worse than other systems. The corresponding audio example is also available online\footnote{\label{demopage}\url{https://bisinger-svs.github.io}}.
\vspace{-0.4cm}
\subsection{Ablation Study}
\vspace{-0.1cm}

We further conduct ablation studies to confirm the effectiveness of the methods proposed in our BiSinger model, which include: 1) using a timbre conversion approach for the monolingual singing dataset, 2) applying pitch shift to the speech dataset, and 3) introducing the Language-style-infused encoder. For clarity, we undertook an additional MOS assessment, where we randomly selected and evaluated 18 distinct test cases based on a single target singer. The results, shown in Table \ref{tab:mos_abl}, confirm the effectiveness of our outlined strategy for multilingual SVS. When compared to \textbf{ASM 1} and \textbf{ASM 2}, \textbf{ASM 3} demonstrates that timbre conversion and the introduction of language ID tokens can improve performance. Moreover, the DB-4 dataset exhibits better performance with the application of pitch shift as opposed to without it.

\begin{table}[H]
\centering
\caption{MOS scores for ablation study.}
\vspace{-0.2cm}

\resizebox{0.41\textwidth}{!}{
\begin{tabular}{lcc|ccc|c}
\toprule
System & SVC & LGT & NUS & DB-4 & pitch-shift & MOS\footnoteref{mos_ci}$\uparrow$ \\
\midrule
\textbf{ASM 1} & -- & -- & $\checkmark$ & -- & -- & $3.83 \pm 0.10$ \\ 
\textbf{ASM 2} & $\checkmark$ & -- & $\checkmark$ & -- & -- & $3.87 \pm 0.10$\\ 
\textbf{ASM 3} & $\checkmark$ & $\checkmark$ & $\checkmark$ & -- & -- & $3.96 \pm 0.09$\\ 
\textbf{ASM 4} & -- & -- & -- & $\checkmark$ & -- & $3.83 \pm 0.10$\\ 
\textbf{ASM 5} & -- & $\checkmark$ & -- & $\checkmark$ & -- & $3.78 \pm 0.10$\\ 
\textbf{ASM 6} & -- & $\checkmark$ & -- & $\checkmark$ & $\checkmark$ & $3.87 \pm 0.10$ \\ 

\bottomrule
\end{tabular}
}
\label{tab:mos_abl}
\end{table}

\section{Conclusion}
\label{sec:conclu}
In this paper, we have developed BiSinger, a system that can synthesize singing voices in both Chinese and English. To overcome language barriers, we adopted a language-independent representation, transitioning from a Pinyin-based annotation to a CMU-based one. Furthermore, we have enhanced the model's performance in English and code-switch SVS by incorporating a monolingual singing dataset using advanced SVC techniques. In addition, we have explored the use of a bilingual speech dataset to facilitate multilingual SVS.  Experimental results demonstrate that our methods can generate multilingual singing voices and enhance English and code-switch SVS while maintaining performance in Chinese songs. In our future work, we will continue to study multilingual and code-switched singing voice synthesis. And it would be promising work to find other solid evaluation metrics for singing voices.
\vspace{-0.3cm}
\section*{\fontsize{10}{12}\selectfont Acknowledgement}
\vspace{-0.3cm}
This research is funded by the Kunshan Municipal Government Research Funding under the project ``Deep Learning based Singing Voice Synthesis for Kun Opera".
\newpage
\pagebreak
\bibliographystyle{IEEEbib}
\bibliography{strings, refs, bisinger}

\begin{thebibliography}{10}

\bibitem{lu2020xiaoicesing}
P.~Lu, J.~Wu, J.~Luan, X.~Tan, and L.~Zhou,
\newblock ``Xiaoicesing: A high-quality and integrated singing voice synthesis
  system,''
\newblock {\em arXiv preprint arXiv:2006.06261}, 2020.

\bibitem{ren2019fastspeech}
Y.~Ren, Y.~Ruan, X.~Tan, T.~Qin, S.~Zhao, Z.~Zhao, and T.-Y. Liu,
\newblock ``Fastspeech: Fast, robust and controllable text to speech,''
\newblock {\em Advances in neural information processing systems}, vol. 32,
  2019.

\bibitem{gu2021bytesing}
Y.~Gu, X.~Yin, Y.~Rao, Y.~Wan, B.~Tang, Y.~Zhang, J.~Chen, Y.~Wang, and Z.~Ma,
\newblock ``Bytesing: A chinese singing voice synthesis system using duration
  allocated encoder-decoder acoustic models and wavernn vocoders,''
\newblock in {\em 2021 12th International Symposium on Chinese Spoken Language
  Processing (ISCSLP)}. IEEE, 2021, pp. 1--5.

\bibitem{wang2017tacotron}
Y.~Wang, R.~Skerry-Ryan, D.~Stanton, Y.~Wu, R.~J. Weiss, N.~Jaitly, Z.~Yang,
  Y.~Xiao, Z.~Chen, S.~Bengio, et~al.,
\newblock ``Tacotron: Towards end-to-end speech synthesis,''
\newblock {\em arXiv preprint arXiv:1703.10135}, 2017.

\bibitem{kim2021conditional}
J.~Kim, J.~Kong, and J.~Son,
\newblock ``Conditional variational autoencoder with adversarial learning for
  end-to-end text-to-speech,''
\newblock in {\em International Conference on Machine Learning}. PMLR, 2021,
  pp. 5530--5540.

\bibitem{zhang2022visinger}
Y.~Zhang, J.~Cong, H.~Xue, L.~Xie, P.~Zhu, and M.~Bi,
\newblock ``Visinger: Variational inference with adversarial learning for
  end-to-end singing voice synthesis,''
\newblock in {\em ICASSP 2022-2022 IEEE International Conference on Acoustics,
  Speech and Signal Processing (ICASSP)}. IEEE, 2022, pp. 7237--7241.

\bibitem{oura2010recent}
K.~Oura, A.~Mase, T.~Yamada, S.~Muto, Y.~Nankaku, and K.~Tokuda,
\newblock ``Recent development of the hmm-based singing voice synthesis
  system—sinsy,''
\newblock in {\em Seventh ISCA Workshop on Speech Synthesis}, 2010.

\bibitem{shi2022muskits}
J.~Shi, S.~Guo, T.~Qian, N.~Huo, T.~Hayashi, Y.~Wu, F.~Xu, X.~Chang, H.~Li,
  P.~Wu, et~al.,
\newblock ``Muskits: an end-to-end music processing toolkit for singing voice
  synthesis,''
\newblock {\em arXiv preprint arXiv:2205.04029}, 2022.

\bibitem{yamamoto2023nnsvs}
R.~Yamamoto, R.~Yoneyama, and T.~Toda,
\newblock ``Nnsvs: A neural network-based singing voice synthesis toolkit,''
\newblock in {\em ICASSP 2023-2023 IEEE International Conference on Acoustics,
  Speech and Signal Processing (ICASSP)}. IEEE, 2023, pp. 1--5.

\bibitem{liu2022diffsinger}
J.~Liu, C.~Li, Y.~Ren, F.~Chen, and Z.~Zhao,
\newblock ``Diffsinger: Singing voice synthesis via shallow diffusion
  mechanism,''
\newblock in {\em Proceedings of the AAAI conference on artificial
  intelligence}, 2022, vol.~36, pp. 11020--11028.

\bibitem{choi2020children}
S.~Choi, W.~Kim, S.~Park, S.~Yong, and J.~Nam,
\newblock ``Children’s song dataset for singing voice research,''
\newblock in {\em International Society for Music Information Retrieval
  Conference (ISMIR)}, 2020.

\bibitem{zhang2022m4singer}
L.~Zhang, R.~Li, S.~Wang, L.~Deng, J.~Liu, Y.~Ren, J.~He, R.~Huang, J.~Zhu,
  X.~Chen, et~al.,
\newblock ``M4singer: A multi-style, multi-singer and musical score provided
  mandarin singing corpus,''
\newblock {\em Advances in Neural Information Processing Systems}, vol. 35, pp.
  6914--6926, 2022.

\bibitem{duan2013nus}
Z.~Duan, H.~Fang, B.~Li, K.~C. Sim, and Y.~Wang,
\newblock ``The nus sung and spoken lyrics corpus: A quantitative comparison of
  singing and speech,''
\newblock in {\em 2013 Asia-Pacific Signal and Information Processing
  Association Annual Summit and Conference}. IEEE, 2013, pp. 1--9.

\bibitem{cai2023cross}
Z.~Cai, Y.~Yang, and M.~Li,
\newblock ``Cross-lingual multi-speaker speech synthesis with limited bilingual
  training data,''
\newblock {\em Computer Speech \& Language}, vol. 77, pp. 101427, 2023.

\bibitem{kenmochi2007vocaloid}
H.~Kenmochi and H.~Ohshita,
\newblock ``Vocaloid-commercial singing synthesizer based on sample
  concatenation.,''
\newblock in {\em Interspeech}, 2007, vol. 2007, pp. 4009--4010.

\bibitem{bonada2016expressive}
J.~Bonada, M.~Umbert~Morist, and M.~Blaauw,
\newblock ``Expressive singing synthesis based on unit selection for the
  singing synthesis challenge 2016,''
\newblock {\em Morgan N, editor. Interspeech 2016; 2016 Sep 8-12; San
  Francisco, CA.[place unknown]: ISCA; 2016. p. 1230-4.}, 2016.

\bibitem{nakamura2014hmm}
K.~Nakamura, K.~Oura, Y.~Nankaku, and K.~Tokuda,
\newblock ``Hmm-based singing voice synthesis and its application to japanese
  and english,''
\newblock in {\em 2014 IEEE International Conference on Acoustics, Speech and
  Signal Processing (ICASSP)}. IEEE, 2014, pp. 265--269.

\bibitem{ren2020deepsinger}
Y.~Ren, X.~Tan, T.~Qin, J.~Luan, Z.~Zhao, and T.-Y. Liu,
\newblock ``Deepsinger: Singing voice synthesis with data mined from the web,''
\newblock in {\em Proceedings of the 26th ACM SIGKDD International Conference
  on Knowledge Discovery \& Data Mining}, 2020, pp. 1979--1989.

\bibitem{zhang2019learning}
Y.~Zhang, R.~J. Weiss, H.~Zen, Y.~Wu, Z.~Chen, R.~Skerry-Ryan, Y.~Jia,
  A.~Rosenberg, and B.~Ramabhadran,
\newblock ``Learning to speak fluently in a foreign language: Multilingual
  speech synthesis and cross-language voice cloning,''
\newblock {\em arXiv preprint arXiv:1907.04448}, 2019.

\bibitem{li2019bytes}
B.~Li, Y.~Zhang, T.~Sainath, Y.~Wu, and W.~Chan,
\newblock ``Bytes are all you need: End-to-end multilingual speech recognition
  and synthesis with bytes,''
\newblock in {\em ICASSP 2019-2019 IEEE International Conference on Acoustics,
  Speech and Signal Processing (ICASSP)}. IEEE, 2019, pp. 5621--5625.

\bibitem{cao2019end}
Y.~Cao, X.~Wu, S.~Liu, J.~Yu, X.~Li, Z.~Wu, X.~Liu, and H.~Meng,
\newblock ``End-to-end code-switched tts with mix of monolingual recordings,''
\newblock in {\em ICASSP 2019-2019 IEEE International Conference on Acoustics,
  Speech and Signal Processing (ICASSP)}. IEEE, 2019, pp. 6935--6939.

\bibitem{nekvinda2020one}
T.~Nekvinda and O.~Du{\v{s}}ek,
\newblock ``One model, many languages: Meta-learning for multilingual
  text-to-speech,''
\newblock {\em arXiv preprint arXiv:2008.00768}, 2020.

\bibitem{liu2019cross}
Z.~Liu and B.~Mak,
\newblock ``Cross-lingual multi-speaker text-to-speech synthesis for voice
  cloning without using parallel corpus for unseen speakers,''
\newblock {\em arXiv preprint arXiv:1911.11601}, 2019.

\bibitem{liu2020multi}
Z.~Liu and B.~Mak,
\newblock ``Multi-lingual multi-speaker text-to-speech synthesis for voice
  cloning with online speaker enrollment.,''
\newblock in {\em Interspeech}, 2020, pp. 2932--2936.

\bibitem{lee2018learning}
Y.~Lee, S.~Shon, and T.~Kim,
\newblock ``Learning pronunciation from a foreign language in speech synthesis
  networks,''
\newblock {\em arXiv preprint arXiv:1811.09364}, 2018.

\bibitem{yang2022improve}
F.~Yang, J.~Luan, and Y.~Wang,
\newblock ``Improve bilingual tts using dynamic language and phonology
  embedding,''
\newblock {\em arXiv preprint arXiv:2212.03435}, 2022.

\bibitem{mcauliffe2017montreal}
M.~McAuliffe, M.~Socolof, S.~Mihuc, M.~Wagner, and M.~Sonderegger,
\newblock ``Montreal forced aligner: Trainable text-speech alignment using
  kaldi.,''
\newblock in {\em Interspeech}, 2017, vol. 2017, pp. 498--502.

\bibitem{morise2016world}
M.~Morise, F.~Yokomori, and K.~Ozawa,
\newblock ``World: a vocoder-based high-quality speech synthesis system for
  real-time applications,''
\newblock {\em IEICE TRANSACTIONS on Information and Systems}, vol. 99, no. 7,
  pp. 1877--1884, 2016.

\bibitem{kong2020hifi}
J.~Kong, J.~Kim, and J.~Bae,
\newblock ``Hifi-gan: Generative adversarial networks for efficient and high
  fidelity speech synthesis,''
\newblock {\em Advances in Neural Information Processing Systems}, vol. 33, pp.
  17022--17033, 2020.

\bibitem{radford2023robust}
A.~Radford, J.~W. Kim, T.~Xu, G.~Brockman, C.~McLeavey, and I.~Sutskever,
\newblock ``Robust speech recognition via large-scale weak supervision,''
\newblock in {\em International Conference on Machine Learning}. PMLR, 2023,
  pp. 28492--28518.

\end{thebibliography}

\end{document}